# Microcell Hot Spot and Smart Antennas Evaluation in WCDMA Macrocell System[(*)]

Carlos H. M. Lima Francisco R. P. Cavalcanti Vicente A. de Sousa Jr. Emanuel B. Rodrigues
carlos@gtel.ufc.br rod@gtel.ufc.br vicente@gtel.ufc.br emanuel@gtel.ufc.br
GTEL-UFC: Wireless Telecom Research Group, Federal University of Ceara, Fortaleza, Brazil.
URL: www.gtel.ufc.br

*Abstract* - **The major purpose of this contribution is to analyze an hierarchical cell structure with a hot spot embedded in a WCDMA macrocellular system. Particular attention is paid to the impact of the introduction of smart antennas in this context. A system-level simulation approach is employed where we consider two main solutions to deal with the hot spot traffic: the placement of a microcell antenna or the adoption of smart antenna technology at the macrocell. Relative gains in signal-to-interference ratio (SIR) are then evaluated. Distinct scenarios are defined based on the proportion of users of different service classes. An evaluation of which solution is better suited for each scenario is then carried out.**

## I. Introduction

This article focuses on a multi-cellular system-level simulation of a WCDMA mobile communication radio access network in the reverse link. The specific case of a macrocell cluster with a Hot Spot (HS) embedded in it is considered. Two main solutions to deal with the additional hot spot traffic are the placement of a microcell antenna near the hot spot or the adoption of smart antenna technology at the macrocell site closest to it. A third approach, interesting for the current investigation, is to combine both solutions to verify if a synergistic effect is obtained.

The remainder of this paper is organized as follows. In section II, we present the model for system-level simulations. The microcell and smart antennas strategies in focus are briefly reviewed in section III. In section IV, we analyze the basic Hierarchical Cell Structure (HCS) arrangement. Macrocell smart antenna performance is presented in section V. The performance results of the combined Microcell and Macrocell Smart Antenna are assessed in section VI. Finally, in section VII, we conclude this contribution with final remarks.

## II. System-Level Simulation Model

### A. General Description

The simulator models a WCDMA system with multiple macrocells with a Hot Spot embedded in one sector of the central cell. Each macrocell site is composed by three hexagonal sectors with radius *R*. A microcell site is comprised of just one hexagonal sector with radius *r* targeted at covering the Hot Spot area. See Fig. 1 for an illustration of the system geometry.

A single-code transmission scheme (based on variable spreading gain - VSG - technique) is employed to provide multi-rate services. We look into two distinct classes of services: speech users and data users. The spatial distribution of the users is uniform in the system coverage area including the Hot Spot area where a higher user density is applied.

We evaluate the interference in the uplink from both macrocell and microcell sites under an hierarchical structure that share the same frequency band taking into account two interferer macrocell layers.

**Figure 1:** Microcell Antenna Geometry.

Our analysis is based on independent trials where mobiles are randomly located obeying the same probability over the whole network. The simulation is semi-static in the sense that there is no user mobility involved and performance is calculated over several uncorrelated location sets.

### B. The Propagation Model

The propagation channel follows the usual path loss/shadowing model where the received power is inversely proportional to the $n^{th}$ power of distance with a spatially correlated log-normal fluctuation. Fast fading is assumed to be averaged out by means of diversity (spatial, Rake reception, etc.). The 2D correlated shadowing model used in this work follows the method presented in [1]. We consider that a shadowing fading sample is composed of two separate parts: the surrounding terrain component and mobile station-base station (MS-BS) path component. The surrounding terrain shadowing component is specific to the location of the user's equipment and the MS-BS path shadowing component is related to the path between each user and each site. The surrounding terrain specific contributions are user location dependent only and the path specific contributions are both user and site location dependent.

For the microcell propagation model, we need to distinguish between line-of-sight (LOS) and non-line-of-sight (NLOS) situations. Thus, a "switching" area centered at the Microcell antenna is employed where the propagation loss exponent is n=2 (areas with LOS). Outside this area, the propagation loss exponent is the same as for macrocells (n=4).

Following this propagation model, the transmitted signal power of the $i^{th}$ MS fitted in the $k^{th}$ service class and connected to the $j^{th}$ BS can be expressed as in Eq. 1.

$$P_{T_{i,j,k}} = \frac{P_{\mathrm{Req}_{j,k}} \cdot r_{i,j}^{\;n}}{10^{\left( x_{\sigma_{i,j}} / 10 \right)} \cdot G_{REF_{i,j}}(\theta_{i,j})} \qquad (1)$$

---

(*) This work was supported by the Ericsson Research - Brazilian Branch under the ERBB/UFC.01 Technical Cooperation Contract. URL: www.ericsson.ufc.br.

where $P_{\mathrm{Req}_{j,k}}$ is the j^th BS received signal power required by the k^th service class; $G_{REF_{i,j}}(\theta_{i,j})$ is the reference system antenna gain of the i^th user with respect to the j^th BS for its $\theta_{i,j}$ direction of arrival (DOA); $r_{i,j}$ is the j^th BS to i^th MS separation distance; $n$ is the path loss exponent; $X_{\sigma_{i,j}}$ is a zero-mean log-normal distributed random variable with standard deviation $\sigma_{dB}$, which describes the shadowing effects for the (i,j) path. In simulations, the reference system uses the sectored antenna radiation pattern of [2].

For the snapshot simulator under study, no dynamic handoffs are modeled. In our simulations a signal strength-based selection combining soft-handover algorithm is employed for cell selection by selecting the smallest $P_{T_{i,j,k}}$ value in Eq. 1. It also affects system performance due to the macro-diversity gain it brings.

Our call admission as well as power control algorithms are based on a signal-strength criterion. A user is blocked if its required transmission power is higher than a threshold $P_{T_{MAX}}$. This threshold is established based on the service class that requires the highest transmission power (higher data rate) when a coverage criterion is to be met at the cell edge. In order to calculate the minimum permitted signal power, we consider a power control dynamic range.

Using a cell edge coverage criterion, it is possible to calculate the transmitted power threshold as shown in Eq. 2. Note that this procedure is performed in the macrocell site because it represents the worst case with respect to the microcell site.

$$P_{T_{MAX}} = P_{\mathrm{Req}_{MAX}} .(2R)^n .10^{\left(\xi/10\right)} \tag{2}$$

where $P_{\mathrm{Req}_{MAX}}$ is the maximum required power among all user service classes, $R$ is the macrocell sector radius and $\xi$ is the shadow fading margin. In our case a 90% boundary coverage criterion is used.

Data is collected from one sector of the central BS (Target Sector) where the SIR is measured taking into account inter and intra-cellular interference. Eq. 3 expresses the interference which the i^th MS connected to the j^th BS causes at the central cell site after power control is performed considering the k^th user service class (see Fig. 2).

$$\mathrm{I}_{i,j,k} = P_{T_{i,j,k}} .\left( r_{i,0}^{-n} .10^{\left(X_{\sigma i,0}/10\right)} . G_{REF_{i,0}}(\theta_{i,0}) \right) 10^{\left(X_{IPC}/10\right)} \tag{3}$$

where $X_{IPC}$ is a zero-mean log-normal distributed random variable with standard deviation $\sigma_{IPC}$, which models imperfect power control.

After smart antennas have been added to the target sector of the central base station, then Eq. 3 is modified to Eq. 4.

$$\mathrm{I}'_{i,j,k} = \mathrm{I}_{i,j,k} .G_{SMART}(\theta_{i,0}) \tag{4}$$

where $G_{SMART}(\theta_{i,0})$ is the smart antenna gain performed on the target sector for the i^th MS connected to the j-th BS.

Spatial channel modeling at system-level is based on the assumption of a planar wave arrival over the azimuth and idealized DOA knowledge, considering solely the mobile position on the cell grid. This scenario can be representative of a system with small angle spread. For a description of robust DOA estimation methods see [3].

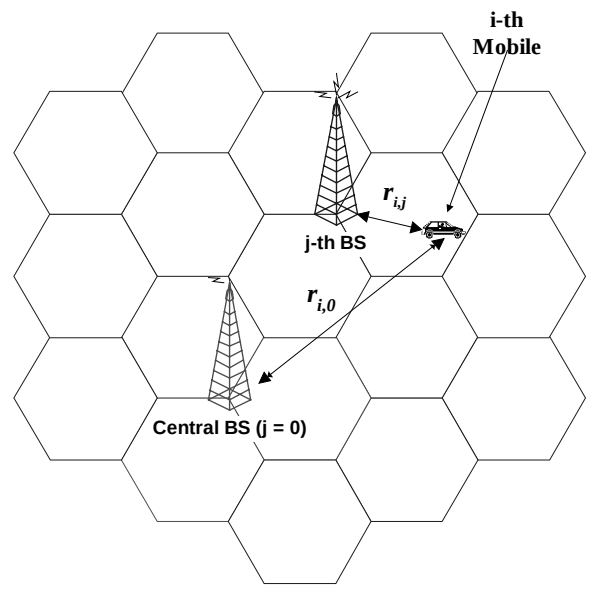

**Figure 2**: Interference Model: The mobile connected to the j^th base station causes inter-cell interference to central base station.

### C. Traffic Model and Power Allocation

The system is assessed considering two distinct service classes: speech service (SPC) at 15 Kbps and high data rate (HDR) at 144 Kbps. The latter may characterize intense data transfer applications such as constant bit-rate videoconference in the uplink. The WCDMA system simulator main parameters are based on the UTRA-FDD specification [4]. The variable processing gain technique is employed as the multi-rate scheme. We neglect additional multi-user or intersymbol interference that may arise from the reduced spreading gain in high bit rate services since advanced techniques such as multi-user detection and equalization can be used in order to avoid such problems [5].

The analysis considers sets of system configurations where the users may be active or not depending on the activity factor. This on-off process is applied independently in each BS.

Our QoS measure is signal-to-interference ratio (SIR) expressed as the bit energy-to-interference density ratio (Eb/Io). Rake receiver, spatial diversity and multi-user detection techniques allow us to estimate a value of SIR = 5dB for SPC users and 7dB for HDR users, as sufficient to provide a satisfactory raw Bit Error Rate (BER) [6]. Further channel coding should modify the BER to the required acceptable level. We use this value of SIR to specify outage probabilities and to optimize power allocation for all service classes, as described in the sequel.

For inter-service class transmission power allocation, we follow the method presented in [7] where power is allocated trying to preserve the prescribed quality of service of each service class. This power allocation scheme is based on the processing gain, required QoS and activity factor of each service class. Eq. 5 gives the power relationship between SPC and HDR service classes.

$$\frac{P_{\mathrm{Req}_{HDR}}}{P_{\mathrm{Req}_{SPC}}} = \frac{SF_{SPC}/\gamma_{SPC} + \alpha_{SPC}}{SF_{HDR}/\gamma_{HDR} + \alpha_{HDR}} \tag{5}$$

where SF is the processing gain, $\gamma$ is the required SIR and $\alpha$ is the activity factor for the appropriate class.

The microcell boundary establishment is the start point to define the microcell power allocation in HCS structure. The boundary between the cell sites (microcell/macrocell) is defined as the equilibrium point, where the required transmission signal power from the user by both cells is the same [8], [9]. Then, considering one user located on the microcell/macrocell border (see point A in Fig. 1), and following our cell boundary definition, considering the average path loss only, we can reach Eq. 6.

$$P_{\mathrm{Req}_{\mu}} = P_{\mathrm{Req}_{M}} \frac{(D-2r)^{n_M}}{(2r)^{n_\mu}} \tag{6}$$

where $n_M$ and $n_\mu$ are the macrocell and microcell path loss exponent, respectively. $P_{\mathrm{Req}_M}$ and $P_{\mathrm{Req}_\mu}$ are the macrocell and microcell required signal power, respectively. The distances *D* and *r* can be checked in Fig. 1.

### *D. Definition of Test Scenarios*

In order to evaluate the different environment, it is necessary to define a few simulation scenarios. In the scenario definition, we follow the suggestions in [10]. Two distinct scenarios are defined: speech scenario considers only SPC users uniformly distributed over the whole cellular network and urban scenario comprises 70% of SPC users and 30% of HDR users uniformly located over the entire system. For all scenarios previously described, we concentrate the same amount of users of a macrocell sector in the Hot Spot embedded in the target central sector. In this situation, the microcell to macrocell user superficial density ratio equals 100 (the hot-spot radius is 1/10$^{th}$ of the macrocell radius). When not mentioned, this value is considered in the simulation campaign.

## III. ANTENNA STRATEGIES

In the present work, we assess two antenna strategies, which are employed for alleviating the Hot Spot traffic: microcell antenna and smart antennas.

The microcell antenna is located at a distance *D* from the central macrocell BS on the microcell site border, in the opposite macrocell antenna pattern direction (see Fig. 1). This orientation of the microcell radiation pattern guarantees a better-fitted microcell boundary corresponding to system requirements [8].

Two smart antenna algorithms are considered: switched fixed beam (FB) and spatial matched filter (MF) [11].

Switched Multiple Fixed Beams (FB) are relatively simple smart antennas and are implemented as a set of multiple narrow beams in pre-established positions. The best beam is selected according to the user position in order to give the highest gain. The pitfall of this architecture occurs when the angular position of the user is in-between two beams where the array gain is lower.

Spatial Matched Filter (MF) steers a beam toward the desired user. The steered beam is generated according to desired user's DOA and does not take into account the position of interferers. Therefore, no null-beamforming is performed since it may be considered unrealistic for CDMA applications with many more users than antenna elements in the array.

## IV. RESULTS FOR MICROCELL ANTENNA IN HCS

The system performance is assessed analyzing the post-despreading SIR behavior of the users located inside the Target Sector.

The system main parameters are: the chip rate: 3.84 Mcps; Spreading Modulation and Symbol Modulation: BPSK; Speech Data Rate (SPC): 15 Kbps; High Data Rate (HDR): 144 Kbps macrocell path loss $n_M = 4$; microcell path loss $n_\mu = 2$; power control dynamic range: 50 dB; macrocell shadowing standard deviation: 8 dB; microcell shadowing standard deviation: 10dB; power control error standard deviation: 2 dB; activity factors are 0.4 for SPC users and 0.7 for HDR users; hot-spot radius is 1/10$^{th}$ of the macrocell radius; Minimum Required SIR are 5 dB for SPC users and 7 dB for HDR users; Processing Gain are 27 and 256 for SPC and HDR users, respectively; HS position: half-way macrocell radius; inter-site shadowing correlation: 50%. The uniform linear antenna array (ULA) has 4 or 8 antennas per sector separated at half wavelength, which is considered to be feasible numbers for implementation. The number of fixed beams was also set to 4 or 8, respectively. Considering these parameters, the required power of the HDR service class is 11.25 dB higher than the SPC class one. On the other hand, the microcell required power is 10.10 dB higher than the macrocell's.

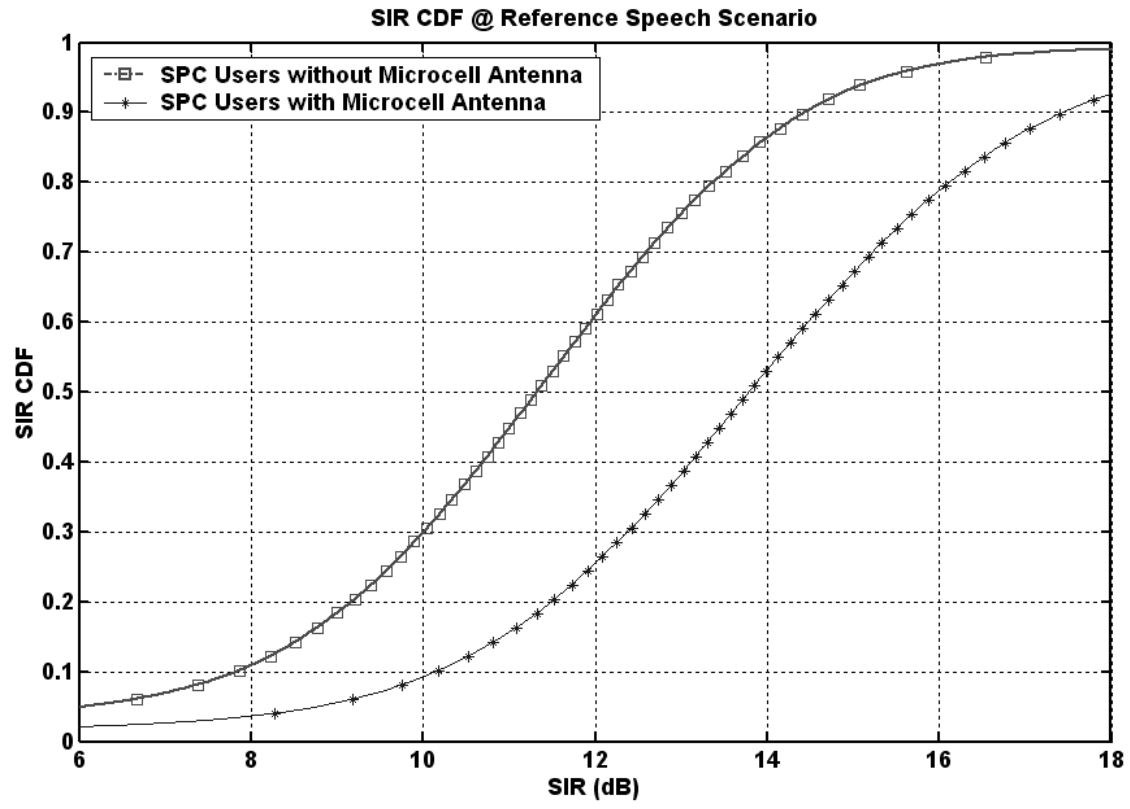

**Figure 3:** SIR CDF before and after the employment of microcell antenna solution in speech scenario (performance of users in the macrocell target sector in HCS observed for the 15 users/sector load).

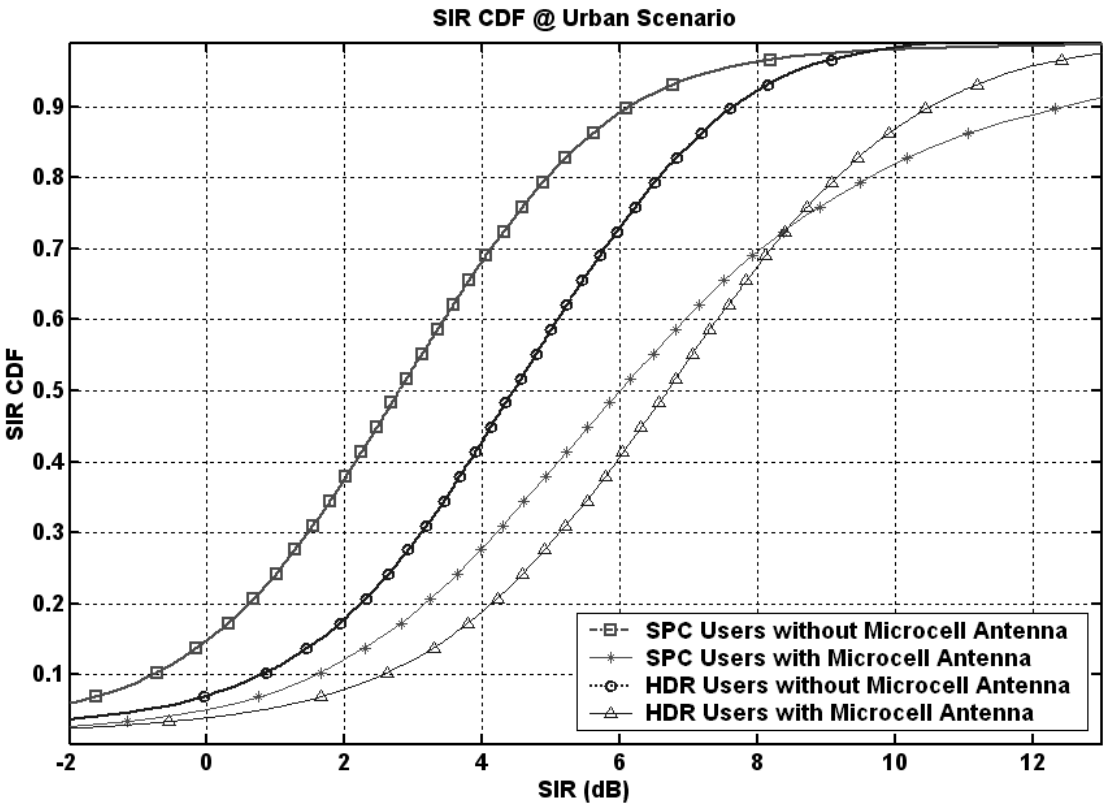

**Figure 4:** SIR CDF before and after the employment of microcell antenna solution in urban scenario (performance of users in the macrocell target sector in HCS observed for the 15 users/sector load).

Fig. 3 shows that the employment of the microcell antenna in the speech scenario provides better SIR CDF. This fact allows us to conclude that this strategy increase the system capacity and improve the user QoS. Fig. 4 draws the SIR CDF performance for users in an urban scenario. This figure presents almost the same microcell antenna gain described for speech scenario, but we can see that by attaching additional services (HDR users), the system performance is degraded. We ascertain that going from speech scenario (I) to urban scenario (II) there is a SIR 10$^{th}$ percentile degradation of approximately 8 dB considering a total load of 15 users per sector.

Since the behavior of SPC and HDR users is similar, we present in the remainder of this section mainly HDR results. For microcell antenna concept in an urban scenario considering users inside and outside the hot spot separately, we observed that the users inside the HS present better gains than those outside the HS. The gains observed by users outside the HS is possible because a great part of the users inside the HS are now connected to the microcell antenna transmitting in a power level lower than that achieved if they were transmitting to the macrocell antenna. Thus, lower interference level is generated. The gain for users inside the HS is due to the higher received power of users connected to the microcell antenna.

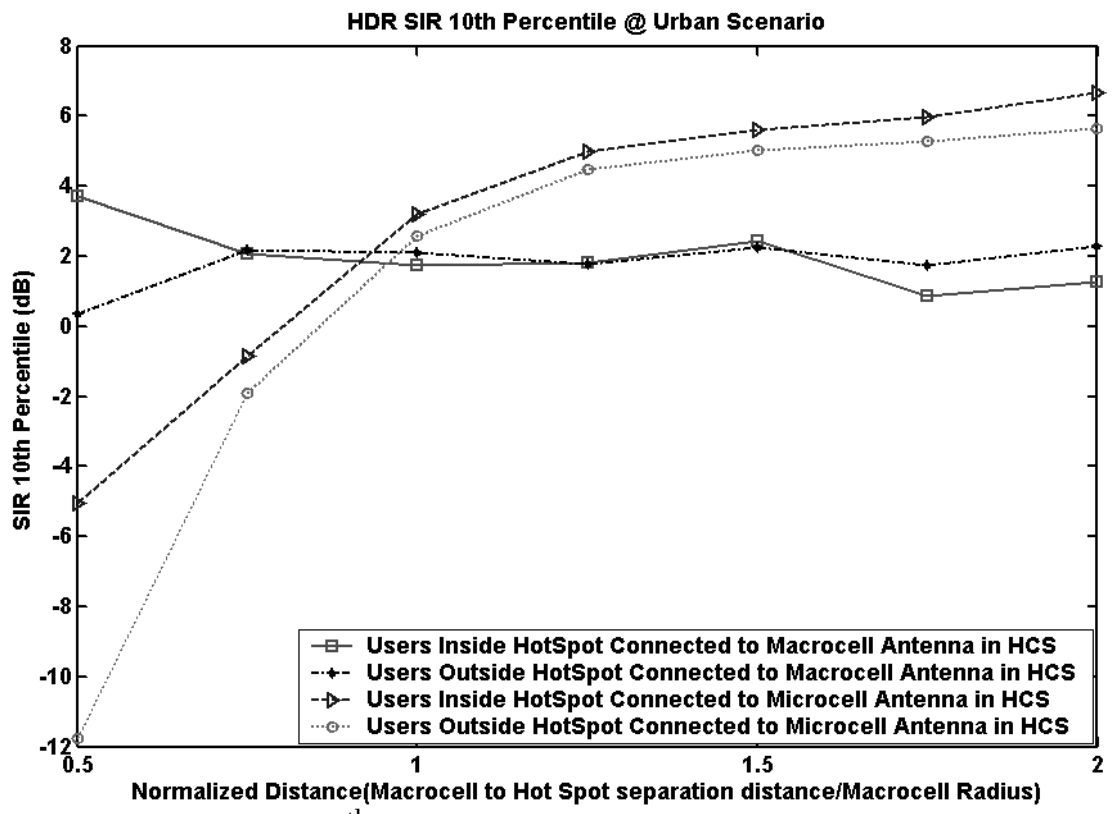

**Figure 5:** SIR 10th Percentile of HDR Users in urban scenario considering HS position displacement from 0.5*R* until 2*R* (*R* - macrocell radius).

Fig. 5 shows the system behavior considering HDR users when changing the HS location. The microcell antenna is positioned on the rightmost corner of the HS demarcation hexagon. Both the hot spot and microcell site are then displaced from half-way the macrocell radius (0.5*R*) to its border (2*R*). Note that the results for $D = R$ correspond to that analysis previously presented. The users connected to the macrocell antenna stay practically unchanged with the displacement. The closer the HS location is from the macrocell hexagonal sector border the higher the gains attained by those users connected to the microcell antenna. That is because the microcell antenna will receive a higher power from its users in order to keep the microcell border (Eq. 6). Another benefit of the HS displacement toward the macrocell border is perceived by users located in this region, since they need to transmit at a lower power to the microcell antenna, causing less interference in the whole system. Moreover, it can be noticed that $D = R$ is a transition point where the microcell users start to experiment higher SIR gains in comparison to the macrocell users. Note that the system is not operational for users connected to the microcell antenna when the distance between the microcell and macrocell sites is lower than R.

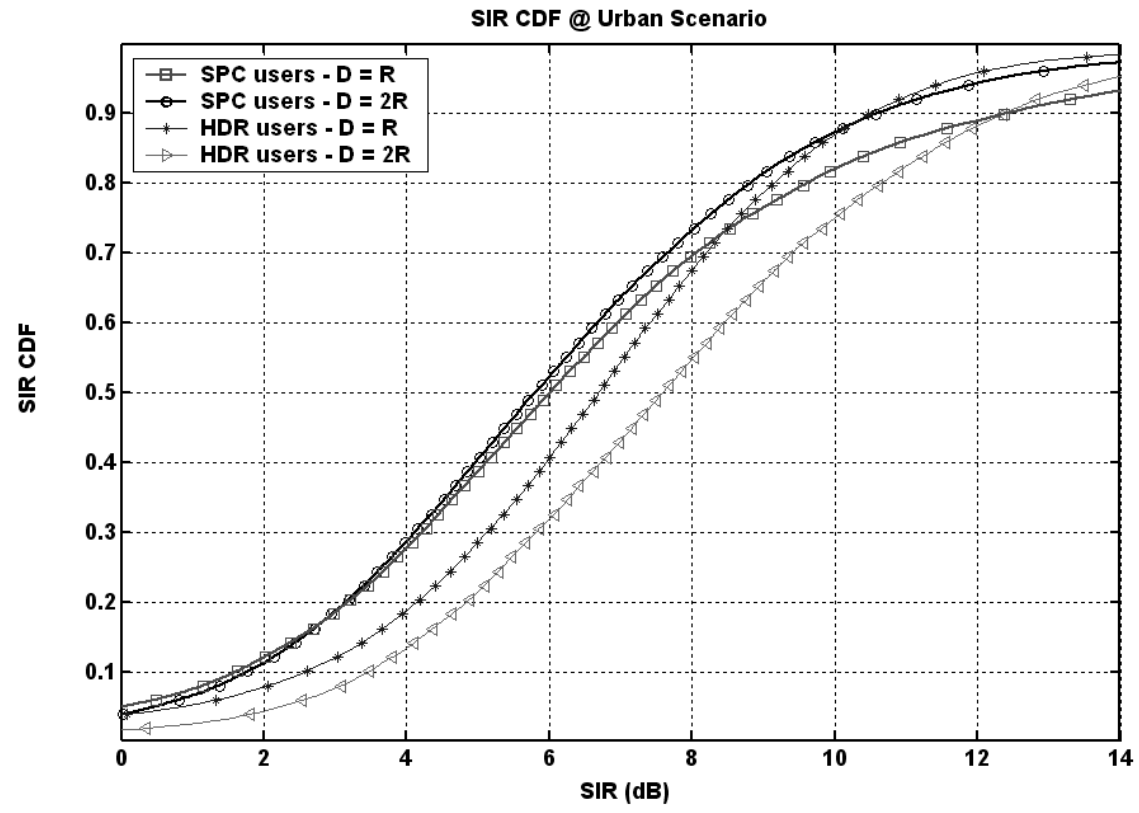

**Figure 6:** SIR CDF of HDR Users in urban scenario with microcell Hot Spot positioned at $D = R$ and $D = 2R$ (performance of users in the macrocell target sector in HCS observed for the 15 users/sector load).

A further analysis of the HS position displacement is presented in Fig. 6. The overall performance of SPC and HDR users (inside and outside the HS) in the form of SIR CDF curves is assessed. It was considered two values for the separation distance between the macrocell and microcell antennas: D = R and D = 2R, where R is the macrocell sector radius. In a general sense, SPC users did not present significant differences in their behavior regarding the change in the HS position. However, HDR users located near the macrocell sector border were beneficiated by the positioning of the HS at D=2R, since they could achieve a better link quality to the microcell antenna rather than to the macrocell one.

Fig. 7 shows the system performance expressed by SIR CDF when the microcell to macrocell user superficial density ratio value increases. We observed that the microcell antenna is not able to cope with the performance degradation caused by the high user superficial density (USD) in the hot spot area. Note that the higher the USD the lower the SIR 10th percentile relative gain between HDR and SPC users. This was expected due to the hostile environment conditions for all users in such a system, regardless of their service class.

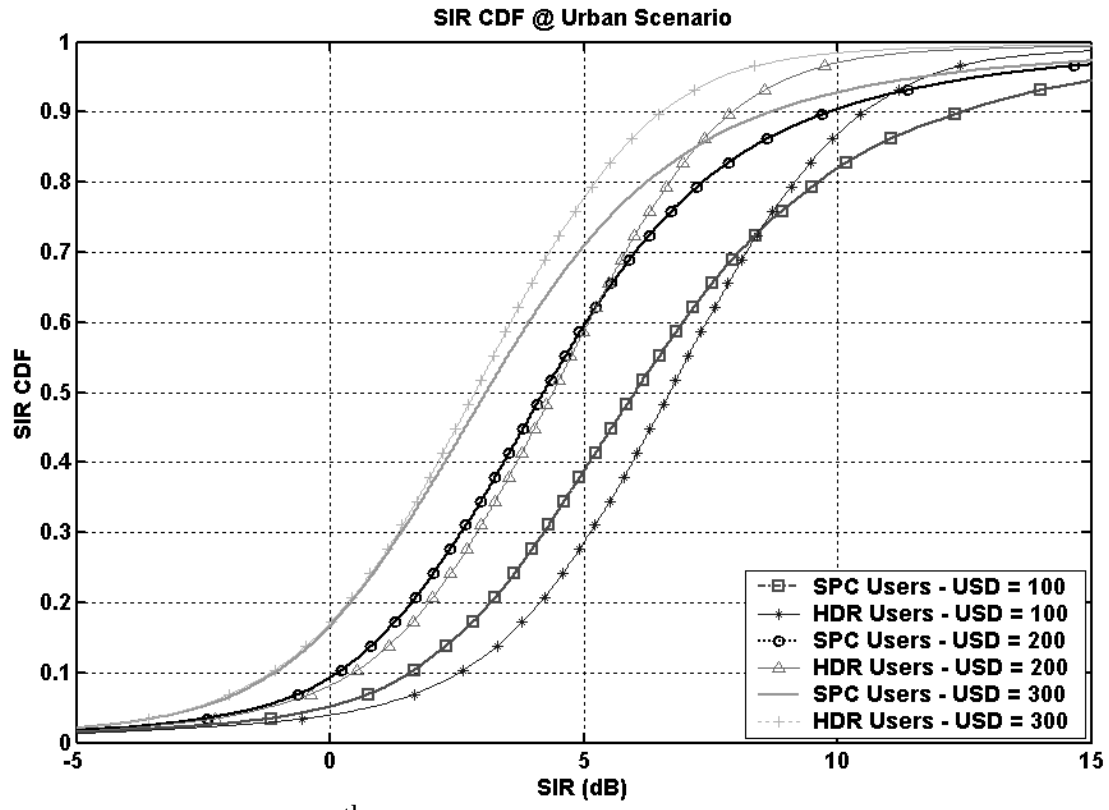

**Figure 7:** SIR 10th Percentile in urban scenario considering the microcell Hot Spot positioned at $D = R$ with different Microcell to Macrocell User Superficial Density Ratio (USD).

## V. Performance Results of the Smart Antennas Strategies at Macrocell Site Only

The following smart antennas (SA) results consider a macrocell antenna array with 4 and 8 antenna elements. Only the interference reduction property of the array is explored and the standard $120^o$ sector pattern is superimposed over the resulting smart antenna pattern.

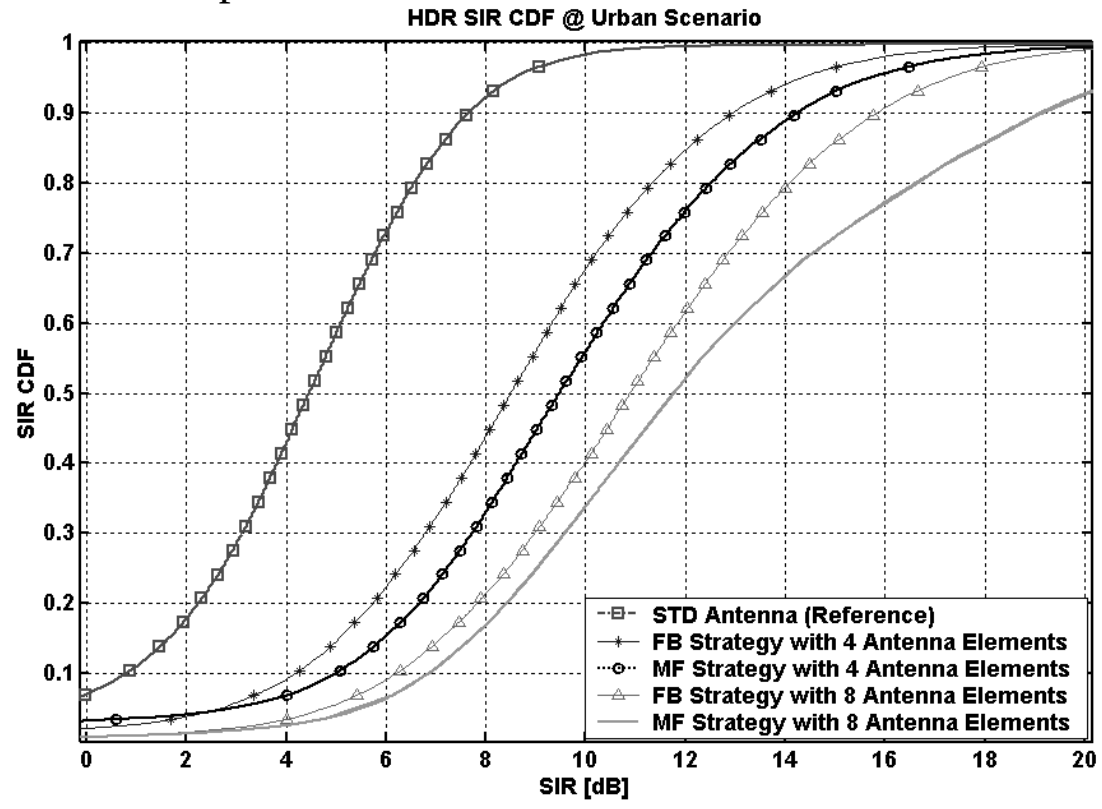

**Figure 8:** HDR SIR CDF before and after the employment of SA solutions in urban scenario without microcell antenna (all users in the target macrocell sector for the 15 users/sector load).

Fig. 8 depicts system fulfillment in urban scenario through analysis of HDR SIR CDF. We can see from this graphic that the overall performance increases considerably after smart antennas approaches have been incorporated. The overall performance corresponds to the joint analysis of users inside and outside the hot spot, namely, all users in the Target Sector. The more sophisticated antenna alternative, in our study MF with 8 antenna elements, the better the performance gains, as is outlined by the rightmost line in the figure.

## VI. Results of Combined Microcell and Macrocell Smart Antennas Strategies

In the last sections, we have presented the performance results of the separated employment of a microcell antenna in HCS environment and the implementation of smart antennas solutions in the macrocell. At this moment, our goal is to assess the system behavior when a combined alternative is performed, namely a combination of microcell and macrocell smart antenna strategies.

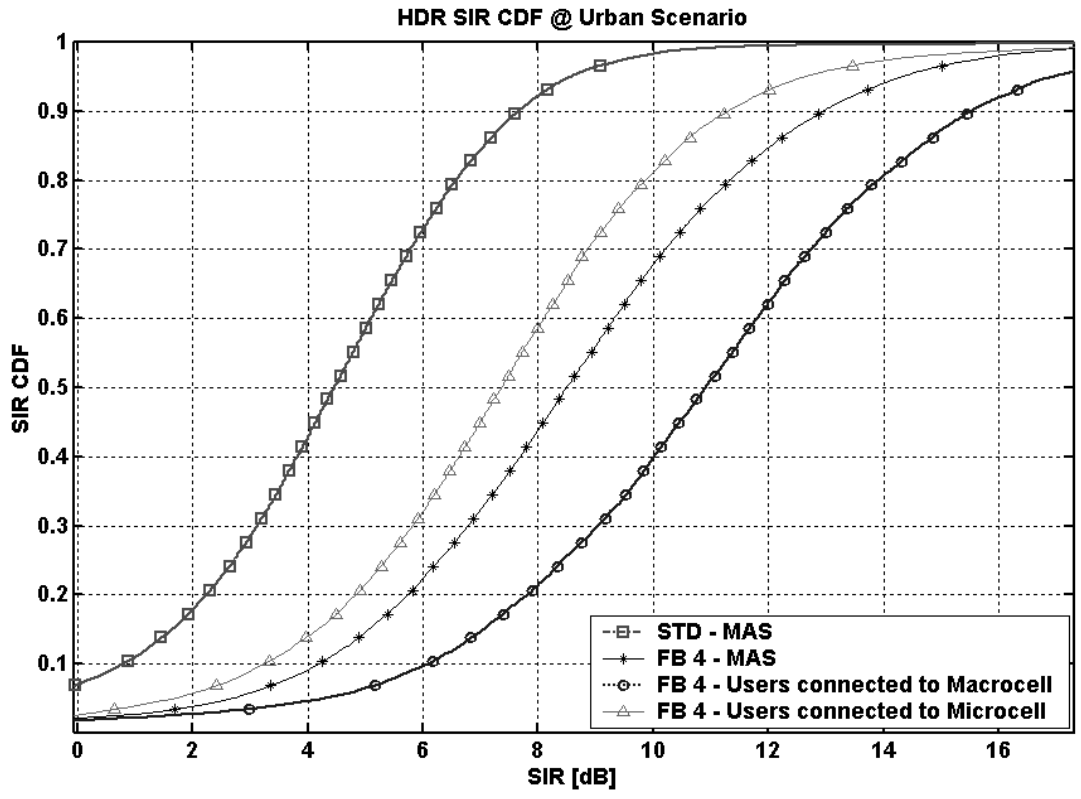

**Figure 9:** HDR SIR CDF before/after the employment of FB smart antenna combined or not with microcell antenna (all users in the target macrocell sector for the 15 users/sector load in urban scenario).

Fig. 9 shows what happens to HDR users at the target sector in urban scenario after implementation of the combined solution. An important characteristic can be observed in this case: the combined solution degrades the overall performance of the users connected to the microcell antenna and increases the SIR of the users connected to the macrocell antenna in comparison to smart antenna in macrocell only. This happened because users connected to the microcell antenna do not benefit from the interference cancellation characteristic of the smart antenna located at the macrocell site. Thus, they observe comparatively higher interference. On other hand, users connected to the macrocell antenna take advantage of the combined solution since the interference offered to the macrocell site was alleviated with the positioning of the microcell antenna in the HS and the interference level observed is also lower due to antenna array processing at the receiver. We observed that fixed beam and matched filter strategies presented similar behavior.

As expected, considering HCS environment as the basis of comparison, we observed that the implementation of any smart antenna solution in the macrocell site only leads to SIR gains of users located outside the HS in both scenarios (speech and urban). Furthermore, users located inside the HS are not influenced by the addition of any smart antenna functionality in the combined alternative.

## VII. Summary of Results

The performance enhancement observed in terms of SIR gains can be interpreted as user or data rate capacity gains, though we have not stated these metrics explicitly in this paper. SIR increase allows more users (or high data rate services) to be admitted in the system as long as the interference does not exceed a specific threshold.

Table I depicts the comparison of the results for all antenna strategies evaluated in this article. As expected, the use of smart antennas enabled higher SIR $10^{th}$ percentile gains, mainly for users located outside the HS. Notice that even for users placed inside the HS, the macrocell smart antenna technique was able to mitigate the external interference. Regarding users inside the HS, smart antennas using an 8-element array performed better than the alternative of 4-element array due to the higher spatial resolution of the former. It can be noticed that in the HCS solution, users inside the hot spot takes advantage of the fact that there is a new and separate antenna for them, but also those users outside HS are beneficiated since the macrocell antenna is less overloaded.

Table I

HDR SIR (dB) $10^{TH}$ Percentile for all antenna alternatives in an Urban scenario for 15 users/sector load; HS located at *D*=*R*; USD = 100 (MAS = macrocell site only).

| | Overall In | Overall Out | Macro In | Macro Out | Micro In | Micro Out |
|---|---|---|---|---|---|---|
| MAS STD | 0,83 | 0,74 | - | - | - | - |
| HCS STD | 3,28 | 1,95 | 1,88 | 1,85 | 3,33 | 2,63 |
| MAS FB4 | 3,66 | 4,92 | - | - | - | - |
| MAS FB8 | 5,65 | 7,08 | - | - | - | - |
| MAS MF4 | 4,46 | 5,95 | - | - | - | - |
| MAS MF8 | 6,06 | 8,57 | - | - | - | - |
| HCS FB4 | 3,44 | 5,09 | 5,66 | 6,11 | 3,36 | 2,53 |
| HCS FB8 | 3,40 | 5,48 | 7,11 | 7,58 | 3,35 | 2,67 |
| HCS MF4 | 3,31 | 5,89 | 7,18 | 7,63 | 3,27 | 2,76 |
| HCS MF8 | 3,27 | 7,11 | 9,49 | 10,86 | 3,18 | 2,71 |

**Overall-In**: performance of users inside the HS in Target Sector; **Overall-Out**: performance of users outside the HS in Target Sector; **Macro-In**: performance of users inside the HS connected to Macrocell Antenna; **Macro-Out**: performance of users outside the HS connected to Macrocell Antenna; **Micro-In**: performance of users inside the HS connected to Microcell Antenna; **Micro-Out**: performance of users outside the HS connected to Microcell Antenna.

The combined alternatives, i. e., microcell antenna and macrocell smart antennas strategies together, did not present the best overall performance because users connected to the microcell antenna do not benefit from the interference cancellation characteristic of the smart antenna located at the macrocell site.

We can conclude that the spatial matched filter with 8-element array implemented at the macrocell site only is the solution which shows the best overall performance in the whole area of the target sector. If one wants to assess the SIR gains distinguishing the user's geographic location and the antenna link connection, HCS MF8 (combined) solution turned out to be the best option for users connected to the macrocell antenna whether they are located inside or outside the HS. Finally, HCS STD, HCS FB and HCS MF solutions showed almost the same performance for users connected to the microcell antenna.

The overall absolute SIR $10^{th}$ percentile gains for all antenna strategies for 5 user/sector load, in order to evaluate the impact of the increase of the user superficial density (USD) within the hot spot area, are depicted in table II. One can observe that in the case of 5 users/sector load and USD=100 (same average amount of users inside and outside the HS), the standard solution (MAS STD) managed to maintain the QoS requirement satisfied (5 dB for SPC and 7 dB for HDR). However, considering the supposition that the average number of users inside the HS was increased by a factor of 3 (USD=300) causing a 2-fold increase in the total number of users in the system, the use of high performance smart antennas is mandatory to guarantee the acceptable user satisfaction level.

TABLE II

OVERALL SIR (dB) 10TH PERCENTILE FOR ALL ANTENNA ALTERNATIVES IN AN URBAN SCENARIO FOR 5 USERS/SECTOR LOAD WITH DIFFERENT MICROCCELL TO MACROCELL USER SUPERFICIAL DENSITY RATIOS (MAS = MACROCELL SITE ONLY).

| | MAS STD | HCS STD | MAS FB4 | MAS FB8 | MAS MF4 | MAS MF8 | HCS FB4 | HCS FB8 | HCS MF4 | HCS MF8 |
|---|---|---|---|---|---|---|---|---|---|---|
| **SPC** USD=100 | 5,03 | 7,35 | 8,14 | 9,65 | 8,80 | 10,13 | 9,25 | 9,64 | 9,51 | 10,03 |
| **HDR** USD=100 | 7,33 | 9,16 | 11,11 | 13,50 | 12,76 | 14,85 | 11,36 | 11,58 | 11,56 | 11,81 |
| **SPC** USD=200 | 3,10 | 5,36 | 5,47 | 6,47 | 5,59 | 6,79 | 6,21 | 6,17 | 6,04 | 6,07 |
| **HDR** USD=200 | 5,11 | 6,12 | 7,42 | 8,76 | 7,88 | 9,13 | 6,43 | 6,28 | 6,24 | 6,26 |
| **SPC** USD=300 | 2,32 | 4,17 | 4,56 | 5,59 | 4,47 | 5,34 | 4,48 | 4,52 | 4,48 | 4,42 |
| **HDR** USD=300 | 4,19 | 4,35 | 6,58 | 7,69 | 6,60 | 7,53 | 4,48 | 4,43 | 4,49 | 4,47 |

Table III shows the overall (inside and outside the HS) SIR 10th percentile for all service classes (SPC and HDR), 3 different average loads (5, 9 and 13 users/sector), 3 distinct microcell to macrocell user superficial density ratios (USD = 100, 200 and 300) and the respective best antenna strategies concerning QoS fulfillment.

It can be seen that the MAS MF8 solution achieved the higher performance (maximum SIR) for low and moderate number of users located inside the HS (USD = 100 and 200). The MAS FB architecture (4 and 8 antenna elements) turned out to be the most flexible solution since it managed to guarantee the minimum QoS requirement in various system demand conditions.

TABLE III

OVERALL SIR 10TH PERCENTILE (dB) FOR THE BEST ANTENNA SOLUTIONS IN AN URBAN SCENARIO FOR DIFFERENT USERS/SECTOR LOADS AND MICROCELL TO MACROCELL USER SUPERFICIAL DENSITY RATIOS (USD)

| | | USD = 100 | | USD = 200 | | USD = 300 | |
|---|---|---|---|---|---|---|---|
| | | **SPC** | **HDR** | **SPC** | **HDR** | **SPC** | **HDR** |
| **5** Users/ Sector | QoSmín | MAS STD | | MAS FB4 | | MAS MF8 | |
| | | 5.03 | 7.33 | 5.47 | 7.42 | 5.34 | 7.53 |
| | SIRmáx | MAS MF8 | | MAS MF8 | | MAS FB8 | |
| | | 10.13 | 14.85 | 6.79 | 9.13 | 5.59 | 7.69 |
| **9** Users/ Sector | QoSmín | MAS FB4 | | - | - | - | - |
| | | 5.36 | 7.41 | | | | |
| | SIRmáx | MAS MF8 | | MAS MF8 | | MAS FB8 | |
| | | 7.60 | 10.25 | 4.56 | 6.48 | 2.81 | 4.46 |
| **13** Users/ Sector | QoSmín | MAS FB8 | | - | - | - | - |
| | | 5.38 | 7.45 | | | | |
| | SIRmáx | MAS MF8 | | MAS MF8 | | MAS FB8 | |
| | | 6.01 | 8.22 | 3.09 | 4.87 | 1.43 | 3.01 |

**QoSmín**: SIR 10th percentile for the least complex antenna alternative that fulfills the signal-to-interference ratio target; **SIRmáx**: SIR 10th percentile for the highest performance antenna alternative;

Furthermore, the MAS FB8 alternative was more satisfactory than the MAS MF8 in the case of USD = 300 (high increase in HS users) and 5 users/sector load. The former was beneficiated since the HS was located in a region of intersection between two fixed beams and the latter did not presented sufficient spatial resolution to resolve this difficult hot spot problem. This fact did not happen when the microcell hot spot was placed at the fixed beam main lobe direction.

Thus, the assessment of the tables presented leads to the conclusion that regarding the complexity and performance trade-off, the MAS FB alternative seems to be the most suitable solution to cope with the hot spot problem.

# VIII. CONCLUSIONS

In a general point of view, the system achieved better performance and higher capacity when the HS was located near the macrocell sector border. Those users who were in a region of poor quality link (macrocell border) were beneficiated with the placement of a microcell antenna near them, thus allowing a lower transmission power and decreasing the whole interference in the system. Considering urban scenario without microcell antenna, we verify that going from $D = R$ to $D = 2R$ there is a SIR 10th percentile improvement of approximately 1 dB considering a total load of 15 users per sector.

We observed that the implementation of any smart antenna solution in the macrocell site only leads to SIR gains of users located outside the HS.

It was observed that the employment of combined smart antenna and microcell solutions degrades the overall performance of the users connected to the microcell antenna and increases the SIR of the users connected to the macrocell antenna in comparison to smart antenna in macrocell only.

We concluded that the system performance in normal conditions for all tested scenarios is improved following the implementation of more sophisticated antennas strategies. Thus, there must be a trade-off between complexity and implementation cost regarding the strategies being assessed.